\documentclass[
 reprint,
 amsmath,amssymb,
 aps,prl
]{revtex4-1}

\usepackage[latin1]{inputenc}
\usepackage[T1]{fontenc}
\usepackage[squaren]{SIunits}
\usepackage{amsmath}
\usepackage{bm}
\usepackage{graphicx}
\usepackage{latexsym}
\usepackage{color}
\usepackage{amssymb}
\usepackage{pdfpages}

\newcommand{\ket}[1]{|#1\rangle}

\newcommand{\ee}{\boldsymbol{\mathcal{E}}}
\newcommand{\eps}{\boldsymbol{\varepsilon}}
\newcommand{\gauss}{\mathrm{G}}
\newcommand{\db}{\mathrm{dB}}
\newcommand{\transm}{\mathcal{T}}

\newcommand{\isol}{\mathcal{I}}
\newcommand{\mean}[1]{\langle #1 \rangle}

\newcommand{\cw}{\circlearrowright}
\newcommand{\ccw}{\circlearrowleft}

\begin{document}

\title{Optical diode based on the chirality of guided photons}

\author{Cl\'ement Sayrin}
\author{ Christian~Junge}
\author{Rudolf~Mitsch}
\author{Bernhard~Albrecht }
\author{Danny~O'Shea }
 \author{Philipp~Schneeweiss} 
\author{J\"urgen~Volz }
\author{Arno~Rauschenbeutel}
 \affiliation{Vienna Center for Quantum Science and Technology, Atominstitut, TU Wien Stadionallee 2, 1020 Vienna, Austria}

%

\date{\today}

\begin{abstract}
Photons are nonchiral particles: their handedness can be both left and right. However, when light is transversely confined, it can locally exhibit a transverse spin\cite{Bliokh12b} whose orientation is fixed by the propagation direction of the photons\cite{Junge13,Neugebauer14,Petersen14,Mitsch14b}. Confined photons thus have chiral character. Here, we employ this to demonstrate nonreciprocal transmission of light at the single-photon level through a silica nanofibre in two experimental schemes. We either use an ensemble of spin-polarised atoms that is weakly coupled to the nanofibre-guided mode or a single spin-polarised atom strongly coupled to the nanofibre via a whispering-gallery-mode resonator. We simultaneously achieve high optical isolation and high forward transmission. Both are controlled by the internal atomic state. The resulting optical diode is the first example of a new class of nonreciprocal nanophotonic devices which exploit the chirality of confined photons\cite{Shen11,Lenferink14,Patent,Xia14} and which are, in principle, suitable for quantum information processing and future quantum optical networks\cite{OBrien07,Kimble08,Ramos14}.
\end{abstract}

\maketitle

\begin{figure}[t]%
\centering
	\includegraphics[width=0.95\columnwidth]{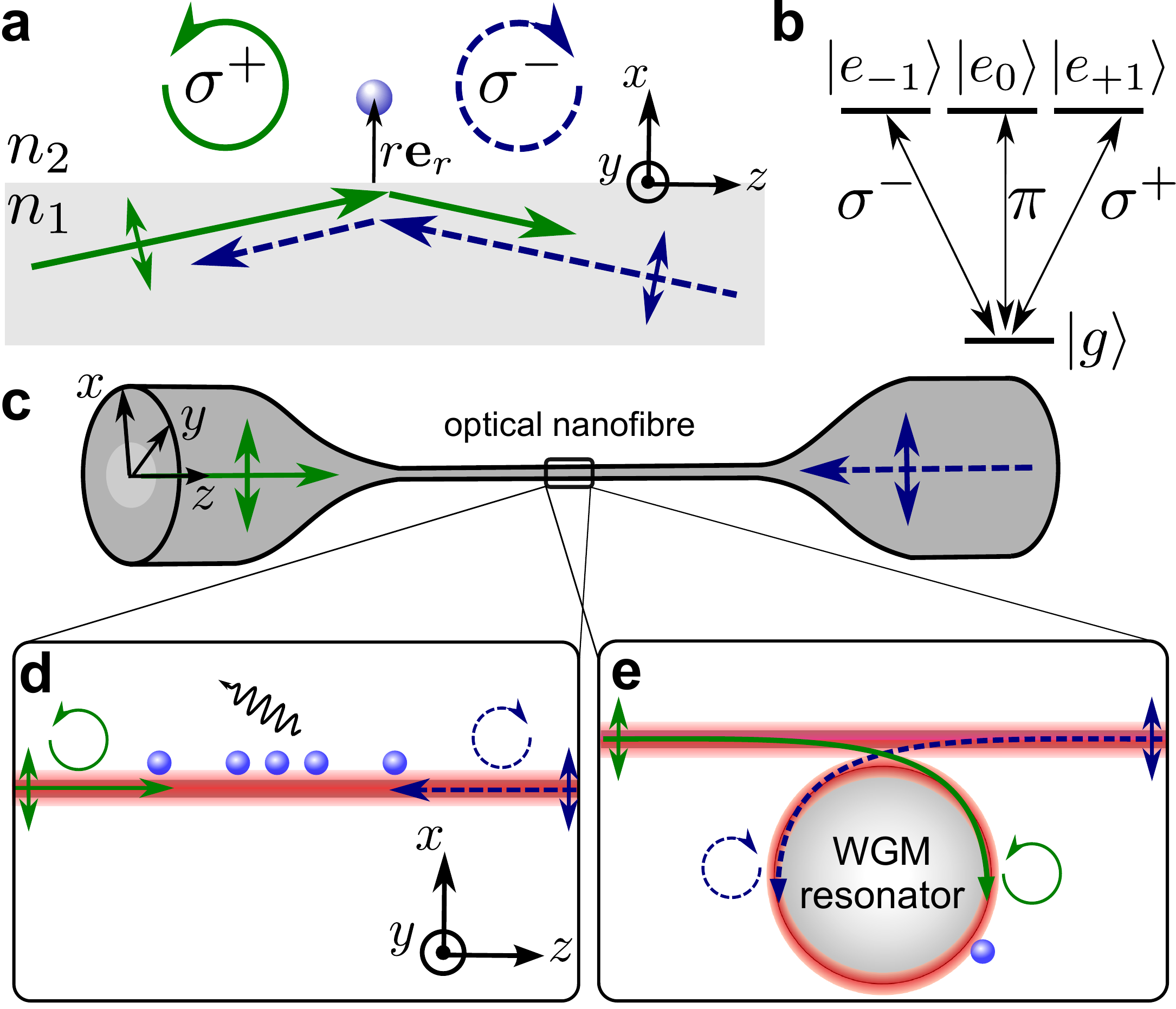}%
	\caption{{\bf Chiral photons in evanescent fields.} {\bf a.} Polarisation properties of the evanescent light field that arises upon total internal reflection of a $p$-polarised wave (polarisation indicated by the double-arrows) at an interface between two dielectric media of refractive indices $n_1>n_2$. Under grazing incidence, an evanescent field that propagates in the $(+z)$-direction is almost fully $\sigma^+$-polarised (green solid arrows). If it propagates in the $(-z)$-direction, it is almost fully $\sigma^-$-polarised (blue dashed arrows). The quantization axis is chosen along $y$, i.e., orthogonal to the propagation direction. An atom (light blue sphere) placed at a distance $r$ to the dielectric interface couples to the evanescent field. {\bf b.} Relevant energy levels of the atom. The ground state $\ket{g}$ is coupled to the excited states $\ket{e_{-1}}$, $\ket{e_0}$, and $\ket{e_{+1}}$ via $\sigma^-$, $\pi$, and $\sigma^+$ transitions, respectively.  {\bf c.} An optical nanofibre is realised as the waist of tapered silica fibre. {\bf d.} Atoms are trapped in the vicinity of the nanofibre, interact with the evanescent field part of the nanofibre-guided modes, and scatter light out of the nanofibre (wavy arrow) with a rate that depends on the direction of propagation. {\bf e.} A single atom is coupled to a whispering-gallery-mode (WGM) bottle microresonator. The atom--resonator coupling strength depends on the propagation direction of the field in the WGM resonator. The WGM resonator is coupled to the optical nanofibre via frustrated total internal reflection.}%
	\label{fig:setup}%
\end{figure}

Light that is transversally confined at the subwavelength scale can exhibit a significant polarisation component along the propagation direction. Such a situation occurs, e.g., in an evanescent field that arises upon total internal reflection of a $p$-polarised wave at a dielectric interface between two media of refractive indices $n_1$ and $n_2<n_1$, see Fig.~\ref{fig:setup}a. Because the longitudinal and transverse components oscillate in phase quadrature, the evanescent field is elliptically polarised. Remarkably, the local ellipticity vector $\eps=i\left(\ee^*\times\ee\right)/{\left|\ee\right|}^2$ is purely transverse\cite{Kawalec07,Bliokh12b}. Here, $\ee$ is the positive-frequency envelope of the electric field. Under grazing incidence, the ellipticity $\varepsilon = |\eps|$  of the the evanescent field reaches its maximum value of\cite{Axelrod84,Kawalec07}
\begin{align}
	\varepsilon_{g} &= \frac{2 \sqrt{1-(n_2/n_1)^2}}{2-(n_2/n_1)^2}.
	\label{eqn:epsilon}
\end{align}
In the case of a silica--vacuum interface ($n_1\approx 1.45$, $n_2=1$), $\varepsilon_g\approx0.95$. As a consequence, when the evanescent field propagates in the $(+z)$-direction, it is almost fully $\sigma^+$-polarised if the $y$-axis is taken as the quantization axis. However, it is almost fully $\sigma^-$-polarised if it propagates in the $(-z)$-direction. This shows that photons in an evanescent field have chiral character: there is an inherent link between their local polarisation and their propagation direction. For plane-waves in free-space, the helicity $\eps\cdot\boldsymbol{\beta}/|\boldsymbol{\beta}|$, where $\boldsymbol{\beta}$ is the wave vector, is equivalent to the chirality. For the case considered in Fig.~\ref{fig:setup}a, however, it does not provide a good measure of the local chirality: Since the local ellipticity vector is transverse, the helicity is zero. Here, we locally define an effective chirality $\chi=\eps\cdot(\boldsymbol{\beta}/|\boldsymbol{\beta}|\times\mathbf{e}_r)$, where $\mathbf{e}_r$ is the normal vector of the interface. With this definition, and in contrast to plane-waves, photons in evanescent fields only have positive chirality, for both the $(+z)$- and $(-z)$-propagation directions.

The chirality of photons in evanescent fields can lead to strongly nonreciprocal behaviour when the photons interact with spin-polarised atoms. The relevant energy levels of the latter are given in Fig.~\ref{fig:setup}b. The light field is assumed to be close to resonance with the transitions from the ground-state $\ket{g}$ to the excited states $\ket{e_{i}}$ where $i\in\{-1,0,+1\}$ denotes the change in magnetic quantum number of the atom with respect to $\ket{g}$. Considering the simplified case where the light is perfectly circularly polarised at the position of the atoms, i.e., $\chi=\varepsilon=1$, the $\ket{g}\to\ket{e_{+1}}$ and $\ket{g}\to\ket{e_{-1}}$ transitions are driven only when the light field propagates in the $(+z)$- and $(-z)$-directions, respectively. The $\ket{g}\to\ket{e_0}$ transition is not driven. 
In order to observe nonreciprocal transmission of light, the atoms should act as polarisation-dependent scatterers, i.e., exhibit different cross-sections for $\sigma^+$- and $\sigma^-$-polarised light. The resulting chiral interaction between each atom and the guided light leads to nonreciprocal transmission as conceptually discussed in Refs.~(\cite{Shen11,Lenferink14,Patent,Xia14}). To realise a polarisation-dependent scattering cross-section, a magnetic field can be applied to lift the degeneracy between the states $\ket{e_{+1}}$ and $\ket{e_{-1}}$. The light can then be made, e.g., resonant with one of the two transitions only. Another strategy consists in choosing $\ket{g}$ such that the two transitions $\ket{g}\to\ket{e_{\pm1}}$ have significantly different strengths. In both cases, the atoms couple unequally to $\sigma^+$- and $\sigma^-$-polarised light. As a consequence, one obtains a strong dependence of the atom--light coupling strength on the propagation direction of the chiral photons. In order to characterise the coupling between the atoms and the light field, we introduce the coefficients $\beta^{(i)} = \beta^{(i)}_++\beta^{(i)}_-$, where
\begin{align}
	\beta^{(i)}_\pm &= \frac{\kappa^{(i)}_\pm}{\kappa^{(i)}_++\kappa^{(i)}_-+\gamma^{(i)}}, \quad i\in\{-1,0,+1\}.
\end{align}
Here, $\kappa^{(i)}_\pm$ is the spontaneous emission rate of an atom in the state $\ket{e_i}$ into the light mode that propagates in the $(\pm z)$-direction and $\gamma^{(i)}$ accounts for all other loss channels such as spontaneous emission into free-space.

We now experimentally demonstrate nonreciprocal transmission of photons that propagate in the evanescent field surrounding a nanophotonic waveguide. We investigate the coupling of these photons with spin-polarised atoms in two qualitatively different regimes: First, an ensemble of atoms interacts with light guided in an optical nanofibre\cite{Vetsch10}. Here, each atom is weakly coupled to the waveguide, i.e., $\kappa_\pm\ll\gamma$ ($\beta\ll1$). Second, a single atom is coupled to an ultra-high quality factor whispering-gallery-mode (WGM) bottle microresonator\cite{Junge13} which, in turn, is critically coupled to the optical nanofibre. This realises an effective strong interaction between the atom and the waveguide, namely $\kappa_\pm\approx\gamma$ ($\beta\approx0.5$). With the atomic ensemble, the ratio of the transmissions in forward and backward direction, i.e. the isolation, is as large as $8~\db$ for a few ten waveguide-coupled atoms, while it is $13~\db$ with the resonator-enhanced scheme. At the same time, the forward transmissions remain as high as $78\,\%$ and $72\,\%$, respectively. Both experiments were carried out in an effective single photon regime, i.e., with a mean number of photons per atomic lifetime much smaller than one. The observed nonreciprocity relies on the chiral character of the guided photons. While nonreciprocal light propagation through integrated optical components has also recently been observed using magneto-optical\cite{Shoji08, Tien11, Bi11}, non-linear\cite{Gallo01,Fan12b, Peng14} and parity-time symmetric materials\cite{Ruter10, Feng11, Regensburger12, Feng13}, and with waveguides with time-modulated properties\cite{Lira12,Tzuang14,ArXiv_Kim14,Estep14}, our work is the first to demonstrate an optical diode that simultaneously provides low optical losses and high optical isolation at the single-photon level.

First, we use an ensemble of individual caesium atoms, see Fig.~\ref{fig:setup}d. The atoms are initially optically pumped towards the state $\ket{g}=\ket{F=4, m_F = +4}$~(\cite{Mitsch14a}), where $F$ is the value of the atomic spin and $m_F$ its projection on the quantization axis. In order to prevent spin flips, a magnetic offset field $B$ is applied along the $y$-axis.
We consider the transmission of a quasi-linearly polarised light field\cite{LeKien14a} that is resonant with the $F=4\to F'=5$ ($\lambda=852$ nm) transition of the caesium D2 line and that propagates through the nanofibre. When its main polarisation is along the $x$-axis, see Fig.~\ref{fig:setup}d, the guided-photons exhibit the required chirality: At the position of the atoms, the local chirality $\chi$ is $0.84$. In the case where the main polarisation axis is along the $y$-axis, the evanescent field is purely linearly polarised at the position of the atoms\cite{Mitsch14b} and $\chi = 0$.

A fraction $(1\pm\chi)/2$ of a light field that propagates in the $(\pm z)$-direction couples to the $\ket{e_{+1}}$ state, while a fraction $(1\mp\chi)/2$ couples to the $\ket{e_{-1}}$ state.
Here, $\ket{e_{+1}} = \ket{F'=5, m_F = +5}$, $\ket{e_0}=\ket{F'=5, m_F=+4}$ and $\ket{e_{-1}} = \ket{F'=5, m_F = +3}$. With this choice, the $\ket{g}\to\ket{e_{+1}}$ transition is 45 times stronger than the $\ket{g}\to\ket{e_{-1}}$ transition.  Furthermore, in this experiment, $B=28~\gauss$, and the laser light is made resonant with the $\ket{g}\to\ket{e_{+1}}$ transition. The $\ket{g}\to\ket{e_{-1}}$ transition is then detuned by $31~\mega\hertz$---a detuning much bigger than the natural line-width of the $\ket{e_{\pm1}}$ states of $2\gamma^{(\pm1)} = 2\pi \times 5.2~\mega\hertz$. Thus, the coupling to the state $\ket{e_{-1}}$ is negligible compared to the coupling to the state $\ket{e_{+1}}$ even when the photons propagate in the $(-z)$-direction. The guided light field is then coupled to an effective two-level atom. In this situation, assuming a weak coherent excitation, the transmission of a light field that propagates in the $(\pm z)$-direction, for a single coupled atom, is given by $\transm_\pm=|t_\pm|^2$, where $t_\pm = 1-2\beta^{(+1)}_\pm$~(\cite{Xia14}). Remarkably, $\beta^{(+1)}_+/\beta^{(+1)}_- = (1+\chi)/(1-\chi)=11.5$ and strongly nonreciprocal transmission through the nanofibre is made possible. 

In our experiment, $\beta^{(+1)}\approx 0.05\ll1$~(\cite{LeKien05c}). This implies that each atom absorbs a few percent of the nanofibre-guided probe light field and that almost all photons that are emitted by the atoms are scattered into free-space. With a single trapped atom, one would therefore expect an isolation $\isol=10\log\left(\transm_-/\transm_+\right) \approx0.3~\db$. In order to enhance the nonreciprocal behaviour of the system, i.e., to increase the isolation, our scheme requires several atoms to interact with the nanofibre-guided light: With $N$ atoms, the transmissions read $\transm_\pm=\left(|t_\pm|^2\right)^N$. The isolation thus scales linearly with $N$.
\begin{figure}%
\centering
\includegraphics[width=0.95\columnwidth]{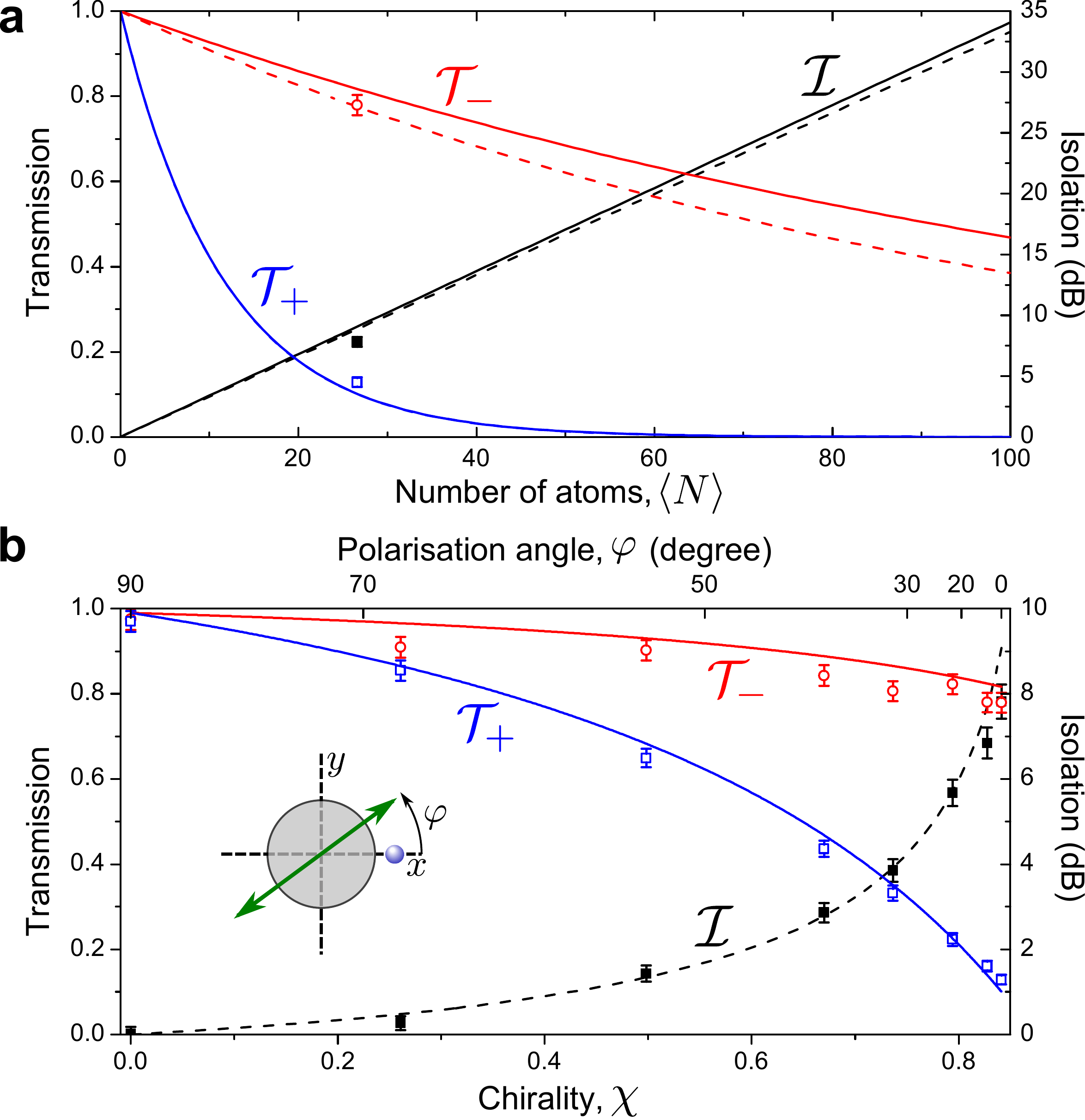}%
\caption{{\bf Nonreciprocal transmission of chiral photons that interact with an ensemble of spin-polarised atoms.} {\bf a.} Isolation $\isol$ (black) and transmissions $\transm_-$ (red) and $\transm_+$ (blue) 
as a function of the mean number $\mean{N}$ of atoms. The simulations are obtained by considering an offset magnetic field of $B = 28~\gauss$ (solid lines) or $B = 0~\gauss$ (dashed lines). The data points correspond to our measurements where we applied a magnetic field of $B=28~\gauss$ and agree with the predictions for $\mean{N} \approx 27$. Each point is the result of the average over 80 experimental runs.
{\bf b.} Transmissions $\transm_-$ (red) and $\transm_+$ (blue) and isolation $\isol$ (black dashed line) as a function of the chirality $\chi$, calculated at the position of the atoms. The lines are the result of a numerical simulation with $\mean{N}\approx27$. Inset: Cross-section of the optical nanofibre (grey disk) including the trapped atoms (blue sphere) and the main polarisation axis of the guided field (green double arrow). The main polarisation axis of the guided field and the $x$-axis enclose the angle $\varphi$. The error bars indicate the 1$\sigma$ statistical error.}
\label{fig:fiberdiode}%
\end{figure}

With a power of the probe light field of $0.8~\pico\watt$, corresponding to $\approx0.1$ photon per excited-state lifetime, we find the transmissions $\transm_+=0.13\pm0.01$ and $\transm_-=0.78\pm0.02$. This strongly nonreciprocal transmission corresponds to an isolation $\isol=7.8~\db$. The residual absorption in the $(-z)$-direction is due to the fact that $\chi<1$.
In Fig.~\ref{fig:fiberdiode}a, we plot, together with our experimental results, the theoretically predicted values of $\transm_+$, $\transm_-$ and $\isol$ as function of the mean number of trapped atoms $\mean{N}$, where $N$ is assumed to follow a Poisson distribution.  Our data agrees very well with the predictions for $\mean{N}\approx27$. Remarkably, with only $\mean{N}\approx90$, one would already obtain an isolation of $\isol = 30~\db$, while the transmission $\transm_-$ would remain as high as $0.47$.  Our scheme thus enables high transmission in the $(-z)$-direction as well as high isolation---two parameters that should be simultaneously maximised for optimal operation of an optical diode.
For comparison, we also plot in Fig.~\ref{fig:fiberdiode}a the values of $\isol$, $\transm_+$ and $\transm_-$ that one would obtain with no magnetic offset field (dashed lines). In this situation, the two $\ket{g}\to\ket{e_{\pm1}}$ transitions are degenerate. The deviation from the situation with $B=28~\gauss$ is small. In the present case, the nonreciprocity thus relies predominantly on the significant difference between the strengths of the $\ket{g}\to\ket{e_{+1}}$ and $\ket{g}\to\ket{e_{-1}}$ transitions while the effect of their detuning is small. 

The local chirality $\chi$ of the photons that interact with the atoms can be continuously tuned via the orientation of the main polarisation axis. The latter is labelled by the angle $\varphi$, see inset of Fig.~\ref{fig:fiberdiode}b. When $\varphi=0$, $\chi$ reaches its maximum of $0.84$, while it vanishes when $\varphi=90\degree$. In Fig.~\ref{fig:fiberdiode}b, we plot the values of $\transm_+$ and $\transm_-$ for different values of $\chi$. Our measurements are in very good agreement with the theoretical predictions calculated with $\mean{N}\approx27$. When $\chi$ is decreased, the isolation decreases. Ultimately, it reaches $\isol=0~\db$ for $\chi=0$, i.e., the transmission becomes symmetric. The evanescent field then couples to the $\ket{g}\to\ket{e_0}$ transition for both propagation directions. Here, because of the applied magnetic field, the light field is out of resonance with the $\ket{g}\to\ket{e_0}$ transition and $\transm_+=\transm_-\approx1$. We can thus control whether the transmission through the nanofibre should be nonreciprocal or reciprocal by choosing $\varphi=0\degree$ or $90\degree$, respectively.

In the above scheme, significant isolation $\isol$ requires many atoms to be trapped in the vicinity of the nanofibre. However, a single atom can be sufficient in the situation where either $\beta_+$ or $\beta_-$ equals $0.5$~(\cite{Xia14}). In this case, the excitation light field and the coherently forward scattered field interfere fully destructively, yielding $\transm_+ = 0$ or $\transm_- = 0$, respectively.
In order to reach the critical coupling condition $\beta_-=0.5$, we need to enhance the atom--waveguide coupling. To this end, we couple a WGM bottle-microresonator\cite{Junge13} to the optical nanofibre, see Fig.\ref{fig:setup}e. The resonator simultaneously provides strong atom--light coupling and strong transverse confinement of light. In the evanescent field, the chirality reaches $\chi=0.94$. A single $^{85}$Rb atom is coupled to the resonator mode, see Methods. Similar to the situation of the nanofibre trapped ensemble, we apply a magnetic field of $B=4.5~\gauss$ along the $y$-axis and prepare the atom in the outermost $m_F=3$ Zeeman substate of the $F=3$ hyperfine ground state. The states $\ket{g}$, $\ket{e_{+1}}$, $\ket{e_0}$ and $\ket{e_{-1}}$ then correspond to the atomic Zeeman states $\ket{F=3, m_F=3}$, $\ket{F'=4, m_F=4}$, $\ket{F'=4, m_F=3}$ and $\ket{F'=4, m_F=2}$, respectively. We tune the resonator and the nanofibre-guided light into resonance with the $\ket{g}\rightarrow \ket{e_{+1}}$ transition. 
Thanks to the different atomic transition strengths, the coupling strengths between the atom and the two counter-rotating resonator modes, $g_{\cw,\ccw}$, strongly depend on the polarisations of the modes and, thus, on their propagation directions. For our experimental settings, we obtain $g_{\cw}=2\pi\times17~\mega\hertz$ ($g_\ccw=2\pi\times 3~\mega\hertz$) for the clockwise (counterclockwise) propagation direction, see Methods.

Here, the combined atom--resonator system plays the role of an effective two-level atom: It possesses two states, namely the clockwise ($i=\cw$) and counter-clockwise ($i=\ccw$) modes of the optical resonator which are dressed by the atom. More precisely, a nanofibre-guided light-field that propagates in the $(+z)$- [$(-z)$-] direction couples solely to the clockwise [counterclockwise] resonator mode with a coupling rate $\kappa$. This means that $\kappa^{(\ccw)}_-=\kappa^{(\cw)}_+=\kappa$ and $\kappa^{(\ccw)}_+=\kappa^{(\cw)}_-=0$. Due to the presence of the atom, the loss rate of the two modes is changed and now reads $\gamma^{(i)} = \gamma_{\rm int} + {g_i}^2/\gamma_{\rm Rb}$~(\cite{Volz14}), where $2\gamma_{\rm Rb}=2\pi\times6\,\mega\hertz$ is the spontaneous decay rate of the $\ket{e_{\pm1}}$ states, $\gamma_{\rm int}=2\pi\times5~\mega\hertz$ is the intrinsic resonator field decay rate and $i\in\{\cw,\ccw\}$. When the light field is resonant with both the bare resonator and the atom, the transmission through the nanofibre is then given as before as
$t_+ = 1-2\beta^{(\cw)}_+$
and
$t_- = 1-2\beta^{(\ccw)}_-$.
Due to its small coupling strength $g_{\ccw}$, the counterclockwise mode is almost unaffected by the coupling to the atoms ($\gamma^{(\ccw)}\approx\gamma_{\rm int}$), while the properties of the clockwise mode are significantly modified. As a consequence, $\beta^{(\cw)}_+\neq\beta^{(\ccw)}_-$ and nonreciprocal transmission should occur.

\begin{figure}[t]
\centering
\includegraphics[width=0.95\columnwidth]{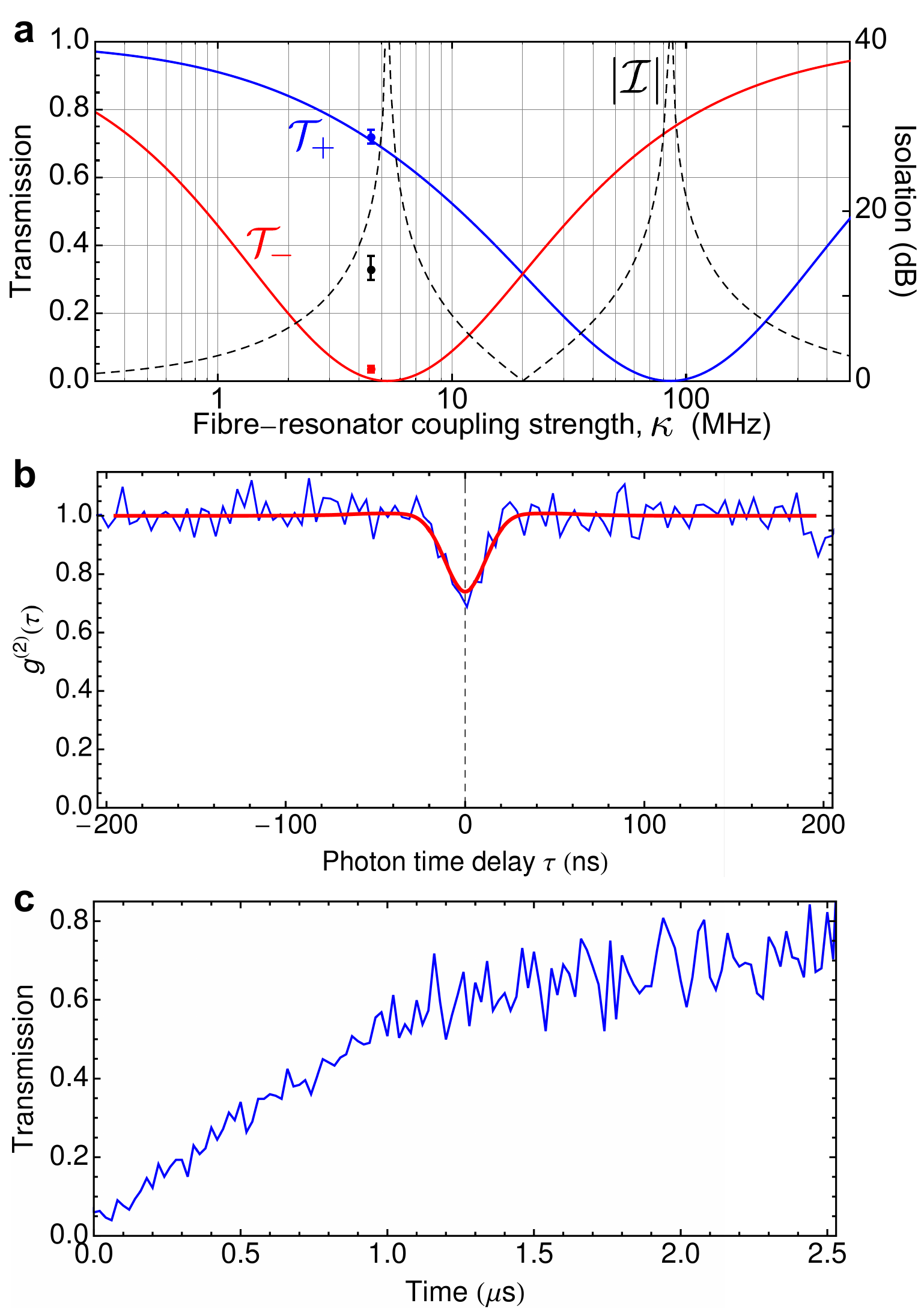}
\caption{{\bf Nonreciprocal transmission of chiral photons that interact with a single resonator-enhanced atom.} {\bf a.} Transmission through the nanofibre in $(+z)$- ($\transm_+$, blue) and $(-z)$-directions ($\transm_-$, red)
and absolute value of the isolation ($|\isol|$, black dashed line) as a function of the fibre-resonator coupling strength $\kappa$.  The solid lines are theoretical calculations and the data points correspond to our measurement. The error bars indicate the 1$\sigma$ statistical error. {\bf b.} Second order correlation function $g^{(2)}$ of the light transmitted through the nanofibre in the ($+z$)-direction. The thin blue (thick red) line corresponds to the measured data (theoretical prediction). For zero time delay one observes photon-antibunching. {\bf c.} Measured transmission $\transm_-$ as function of time. Due to optical pumping into the Zeeman ground state $\ket{F=3,m_F=-3}$, the system changes its transmission properties and, after about $2~\micro\second$, the directionality of the optical diode  reverses.}
\label{fig:wgm}
\end{figure}

To reach $\beta^{(\ccw)}_-=0.5$, one adjusts the resonator--fibre coupling rate $\kappa$ such that $\kappa = \kappa_{\textrm{crit}} \equiv \gamma^{(\ccw)}$. 
In the experiment, we set the resonator--nanofibre coupling strength to $\kappa=\gamma_{\rm int}=2\pi\times5~\mega\hertz$, which corresponds to critical coupling to the empty resonator and is close to the optimal value of $\kappa_{\textrm{crit}}\approx2\pi\times5.5~\mega\hertz$. Here, the power of the probe light field is 3 pW, corresponding to $\approx 0.2$ photon per resonator lifetime. 
We measure a transmission through the nanofibre of $\transm_+=0.72\pm0.02$ in the $(+z)$-direction and of $\transm_-=0.03\pm0.01$ in the $(-z)$-direction. This corresponds to an isolation of $\isol=13~\db$. Figure~\ref{fig:wgm}a shows the measured values together with the theoretical predictions for our system as a function of the fibre--resonator coupling $\kappa$.

In contrast to the ensemble-based scheme, the resonator-based approach relies on a single atom only. As a consequence, one expects a nonlinear response already when only two photons arrive simultaneously\cite{Volz14}. This nonlinearity is apparent in the measurement of the second order correlation function of the light transmitted through the nanofibre in the $(+z)$-direction, see Fig.~\ref{fig:wgm}b. We observe photon anti-bunching which demonstrates that the transmission probability of two simultaneously arriving photons is smaller than for individual photons. The atom-resonator system thus constitutes a nonlinear isolator at the single photon level, where the transmission through the fibre strongly depends on the number of photons in the waveguide. 

Another important aspect of our system is that the nonreciprocal transmission properties depend on the internal state of the atom. In order to show this, we measured the transmission through the nanofibre in the $(-z)$-direction as a function of time, see Fig.~\ref{fig:wgm}c. In this case, the light which enters the resonator is nearly fully $\sigma^-$-polarised. It then drives $\Delta m_F=-1$-transitions and eventually pumps the atom to the ground state $\ket{F=3,m_F=-3}$. For this atomic state, the directional behaviour of the system is reversed: We now observe a high transmission through the waveguide in the $(-z)$-direction. The strength and direction of the optical isolation can thus be controlled via the spin of the atom.

Instead of operating the nonreciprocal waveguide in the dissipative regime, it would also be possible to realise dispersive nonreciprocal elements. This should allow one to implement quantum protocols in which the nonreciprocity is used to generate entanglement\cite{Stannigel12,Ramos14} and, in the case of a nonlinear chiral interaction, to process quantum information that is encoded in single photons\cite{ArXiv_Soellner14}. Chirality of photons occurs in general for all kinds of tightly confined propagating light fields as encountered, e.g., in strongly focused laser beams, photonic crystals, or lithographically fabricated semiconductor waveguides. Moreover, other emitters like quantum dots, defect centres in diamond, and possibly even tailored plasmonic structures can also show the required spin-dependent coupling to light. In view of the generality of the demonstrated effect, we therefore expect chiral light-matter interaction to lead to novel functionalities and components not only in communication and information processing schemes but also for sensing, micro-manipulation, and high-resolution imaging.

\bigskip

\section*{Methods} 
\paragraph{Optical nanofibre} The silica optical nanofibre is realised as the waist of a tapered optical fibre. This enables close-to-unity coupling efficiency of light from the unprocessed fibre to the nanofibre waist and vice versa. The nanofibre has a nominal radius of $a = 250~\nano\meter$ which is small enough so that the fibre only guides the fundamental $\textrm{HE}_{11}$ mode for all wavelengths involved in the experiments.

\paragraph{Ensemble experiment} 
The atoms are located in a nanofibre-based two-colour optical dipole trap\cite{Vetsch10}. Two diametric linear arrays of trapping sites are created by sending a 1064-nm-wavelength red-detuned standing wave, with $0.77~\milli\watt$ per beam, and a blue-detuned running wave with a 783-nm-wavelength and a power of $8.5~\milli\watt$ through the nanofibre. The period of the array is about $0.5~ \micro\meter$, the sites are located $230~\nano\meter$ from the nanofibre surface and they contain at most one atom\cite{Vetsch10}. In this work, the atoms are prepared in only one of the two linear arrays\cite{Mitsch14a}, namely in the one located at $x > a$, see Fig.~\ref{fig:setup}.d.

In order to measure the transmissions $\transm_{\pm}$, two single-photon-counting modules (SPCM) are placed at the two ends of the nanofibre. The number of transmitted photons is measured during $300~\micro\second$, with and without trapped atoms, and is corrected for background counts. The transmissions $\transm_{\pm}$ are given by the ratio of these two numbers.

The theoretical predictions plotted in Fig.~\ref{fig:fiberdiode} are the result of numerical calculations. The latter are based on the model developed in Ref.~(\cite{LeKien14a}). The transmissions $\transm_{\pm}$ are calculated for a single atom ($N=1$) prepared in the $\ket{F=4, m_F=4}$ state, located $230\,\nano\meter$ from the surface of $a=250\,\nano\meter$-radius nanofibre. The model also takes into account the value of the offset magnetic field and the orientation of the main polarisation of the guided light field. The transmissions corresponding to a mean number of trapped atoms of $\mean{N}$ are given by
\begin{align}
\transm_{\pm}(\mean{N}) &= \sum_N p(N,\mean{N}) \transm_{\pm}(N=1)^N,
\end{align}
where $p(N,\mean{N})$ is the Poisson distribution of average $\mean{N}$. The mean number of atoms $\mean{N}\approx27$ used in Fig.~\ref{fig:fiberdiode} is the result of a fit to the experimental data.

\paragraph{Resonator experiment} The bottle microresonator is a highly prolate shaped WGM-resonator with a diameter of 36 $\micro$m fabricated from a standard optical fibre. Its coupling $\kappa$ to the optical nanofibre is tuned by changing the distance between the nanofibre and the resonator surface. 

In the experiment, an atomic fountain delivers a cloud of laser-cooled $^{85}$Rb atoms to the resonator. In order to detect in real time the presence of a single atom in the resonator mode, we critically couple the optical nanofibre and the resonator. An SPCM records the transmission through the nanofibre of a probe light field that is resonant with the empty resonator mode\cite{Junge13}. When an atom enters the resonator mode, the transmission increases by two orders of magnitude and the interaction with the resonator light field optically pumps the atom into the $\ket{F = 3,m_F = 3}$ hyperfine ground state. Using a field programmable gate array-based real-time detection and control system, we react to the increasing count rate within approximately $150\,\nano\second$. Subsequently, we execute our measurement sequence during which we send a light pulse through the nanofibre for a predetermined time interval ranging from $100\,\nano\second$ to a few microseconds along the $(+z)$- or $(-z)$-direction. A final 1-$\micro\second$-probing interval ensures that the atom is still coupled to the resonator mode at the end of the measurement sequence.

\paragraph{Atom--resonator interaction}
In order to obtain the theoretical predictions for the atom--resonator system shown in Fig.~\ref{fig:wgm}, we use the Janyes-Cummings Hamiltonian that describes the interaction of a two-level atom with an optical mode. We extend this model to include the two counter-propagating resonator modes as well as their chiral character and the full Zeeman substructure of the atom. The only free parameter is the average coupling strength $g_\cw$ between a single atom and the resonator mode that propagates in the clockwise direction. From a measurement of the vacuum-Rabi splitting\cite{Junge13}, we obtain $g_\cw=2\pi\times17\,\mega\hertz$. 
Similar to the ensemble case, the coupling strength $g_\ccw$ to the counterclockwise mode is dominated by the residual coupling of the light field to the state $\ket{e_{+1}}$. Given the chirality $\chi=0.94$ of the photons in the evanescent field of the resonator, we obtain $g_\ccw=2\pi\times3\,\mega\hertz$.

To model the nonlinear response of our atom--resonator system and to obtain the $g^{(2)}$-function plotted in Fig.~\ref{fig:wgm}b, we consider the simplified case of a two-level atom that interacts with a single resonator mode. We numerically solve the master equation of the pumped atom--resonator system for the steady state. In particular, we take into account the distribution of the atom--resonator coupling strengths\cite{Volz14}.
 
\bigskip

\noindent $^\ast$To whom correspondence should be addressed:\\
E-mail: schneeweiss@ati.ac.at\\
E-mail: jvolz@ati.ac.at\\
E-mail: arno.rauschenbeutel@ati.ac.at

\smallskip
\noindent $^\dagger$ now at Zernike Institute for Advanced Materials, University of Groningen, Netherlands

\section*{Acknowledgements}
\noindent Financial support by the Austrian Science Fund (FWF, SFB NextLite Project No.~F 4908-N23, SFB FoQuS Project No.~F~4017 and DK CoQuS Project No.~W~1210-N16) and the European Commission (IP SIQS, No.~600645) is gratefully acknowledged. C.S. and J.V. acknowledge support by the European Commission (Marie Curie IEF Grant 328545 and 300392, respectively). C.J. acknowledges support from the German National Academic Foundation.

\section*{Competing Interests}
\noindent The authors declare that they have no competing financial interests.

\section*{Author Contributions}
\noindent For the experiment with the nanofibre-trapped ensemble of atoms: C.S., R.M. and B.A. performed the experiment, and C.S., R.M. and P.S. analysed the data. For the experiment with the WGM-resonator: C.J. and D.O. performed the experiment and J.V. analysed the data. C.S, P.S., J.V. and A.R. wrote the manuscript. All the authors reviewed the manuscript.

\end{document}